\documentclass[hyphens]{svproc}

\usepackage{url}

\usepackage{longtable}
\usepackage{booktabs}   
\usepackage{subcaption} 
\usepackage[unicode=true,colorlinks=false,linkcolor=black,hidelinks]{hyperref}
\usepackage{graphicx}

\usepackage{url}

\IfFileExists{footnote.sty}{\usepackage{footnote}\makesavenoteenv{longtable}}{}
\providecommand{\tightlist}{%
  \setlength{\itemsep}{0pt}\setlength{\parskip}{0pt}}

\usepackage{fancyvrb}

\DefineVerbatimEnvironment{Highlighting}{Verbatim}{commandchars=\\\{\}}

\newlength{\cslhangindent}
\newenvironment{cslreferences}%
  {\setlength{\parindent}{0pt}%
  \everypar{\setlength{\hangindent}{\cslhangindent}}\ignorespaces}%
  {\par}

\begin{document}
\mainmatter

\title{Infer XPath}

\providecommand{\subtitle}[1]{}
\subtitle{Application -- Work in
progress report}

\author{ Michał J.
Gajda \inst{1} \and  Hai Nguyen
Quang \inst{2} \and  Do Ngoc
Khanh \inst{2} \and  Vuong Hai
Thanh \inst{2}}

\institute{
Migamake Pte Ltd\\
\email{migamake@migamake.com}\\
\texttt{}\\  \and Kikai Tech\\
\email{migamake@migamake.com}\\
\texttt{}\\ }

\maketitle

\hypertarget{introduction}{%
\section{Introduction}\label{introduction}}

Semistructured data{[}9{]}
schematisation{[}8{]} has a long history
of research and is one of the most
practical AI challenges{[}7, 10{]}.

Nowadays, with most data available on
the web, it got a new name of
``scraping'' and with it much tarnished
reputation{[}3{]}.

We look at the problem as a discovery of
the answer sets desired by the user.
Coming from XPath selectors{[}2{]}
background we propose a way to construct
selectors automatically, and discover
schemas of the semistructured data by a
combination of lazy path reconstruction,
and lazy preference functions.

\hypertarget{overview}{%
\section{Overview}\label{overview}}

We first offer to expand declarative
language XPath{[}2{]} that was
originally invented for generic XML
documents into XPath BE (Browser
Extensions) with additional operators
that facilitate navigation consistent
with the visual and logical structure of
the web page. We propose new axes that
explicitly navigate the tables and
visual relations between different page
elements.

We also propose how to facilitate
understanding of entire websites, by
providing XPath function that allows for
lazy navigation over the whole website,
instead of just a single page.

As a second step, we to propose a way to
recognise semantic elements on the web
page automatically by extended regular
expression matching on the text content
entire subtrees, instead of nodes as it
is done in plain XPath. This allows us
for fast classification of sets of
semantic values\footnote{By semantic
  value set we indicate a set of
  recognisable entities of the same type
  that can be extracted as text content
  of any subtree within the document(s).}
within the document(s).

Our next step is to propose a
meta-language InferXPath that takes sets
of semantic values and finds possible
queries that connect these sets. Since
there are many possible queries, but we
are only interested in the simplest
query that satisfies our conditions, we
can compute them lazily.

That should give us a set of tables that
provide a structured view of the data
for systematic data analysis.

\begin{figure}[h]

\includegraphics[width=1\textwidth,height=0.4\textheight]{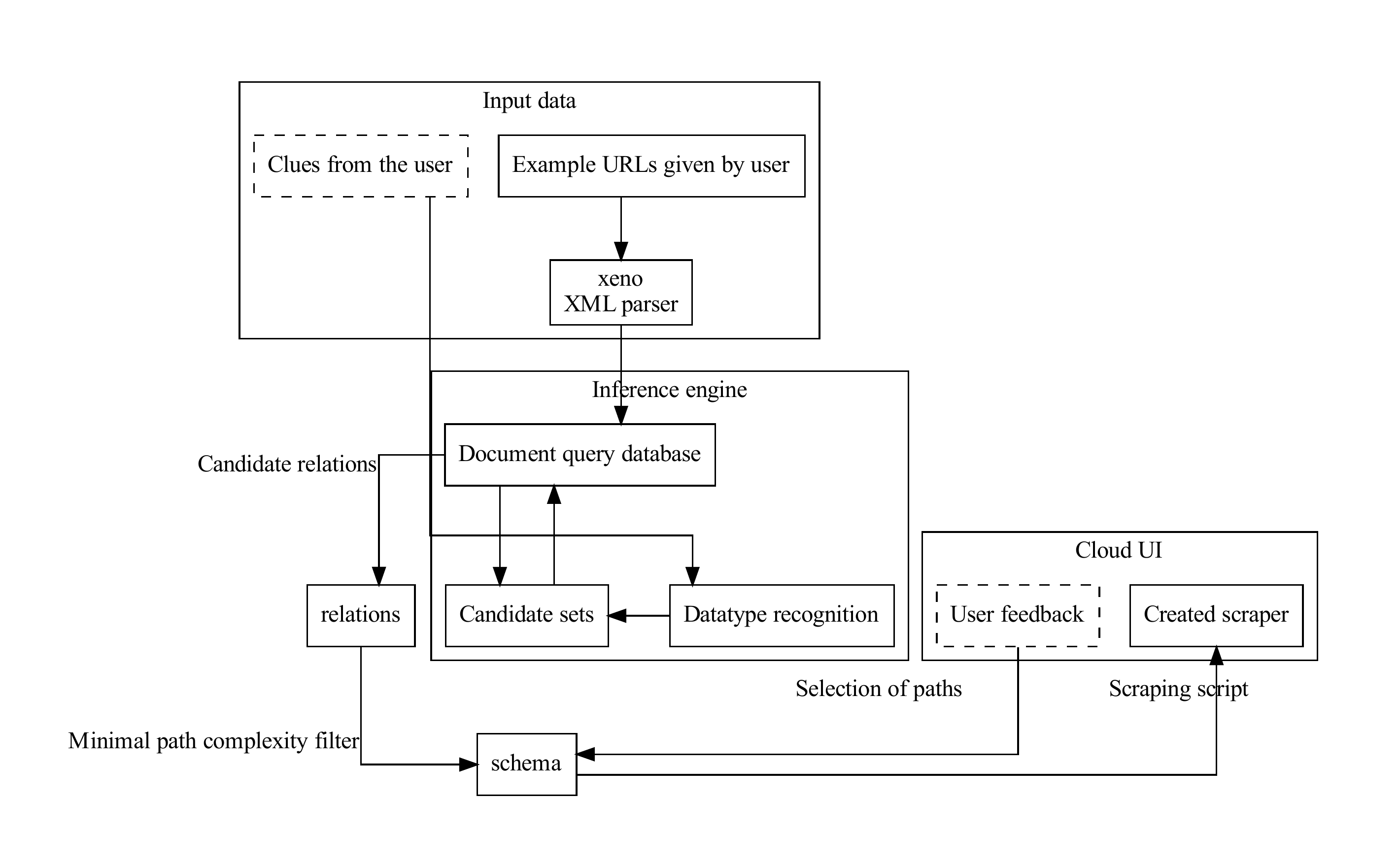}

\caption{Overview of the InferXPath system}
\end{figure}

\hypertarget{xpath-browser-extensions}{%
\section{XPath browser
extensions}\label{xpath-browser-extensions}}

\hypertarget{introduction-to-xpath}{%
\subsection{Introduction to
XPath}\label{introduction-to-xpath}}

XPath is a query language proposed for
the XML documents, and quite popular for
selecting fragments of HTML DOM object
model of the web page. It is a sequence
of selections joined by \texttt{/}
sequencing operator, or
\texttt{\textbar{}} alternative
operator. Each selection is either a
single step across the \emph{axis}, a
function call, or a predicate. Axis step
\texttt{child::*} refers to all the
child elements, or \texttt{root::*}
which moves to a single root element
within the document. Each axis step may
additionally have a condition on the
name of the node selected:
\texttt{child::h1} will only select
\texttt{h1} element node that is a
direct descendant of any node within the
current query set.

Predicates are written as
\texttt{{[}..subquery..{]}} select a set
of nodes that answer the subquery with a
truthy value. Function calls take a
number of arguments that each contains a
node-set described by a subquery, for
example \texttt{concat(a/text())}.

Beside core XPath, there are several
syntactic sugar shorthands for the core
XPath expressions.

It has been frequently proposed to
describe XPath selection expressions by
non-deterministic tree automata
accepting nodes within the selected
node-set. It bears no relation to our
work since we can also precompute
containment and subtree relations as a
part of a document database.

\begin{figure}[h]

\includegraphics[width=0.35\textwidth,height=0.15\textheight]{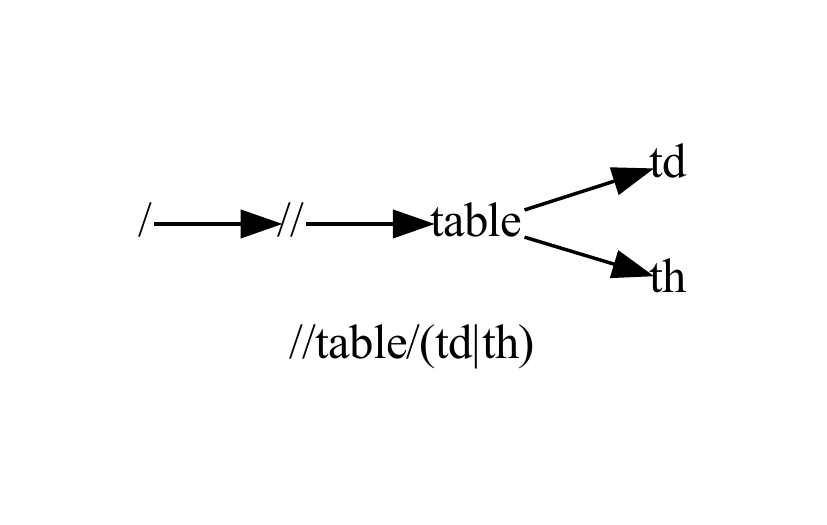}

\caption{Example XPath expression `//table/td|th` rendered as a graph.}
\end{figure}

\begin{figure}[h]

\includegraphics[width=0.45\textwidth,height=0.1\textheight]{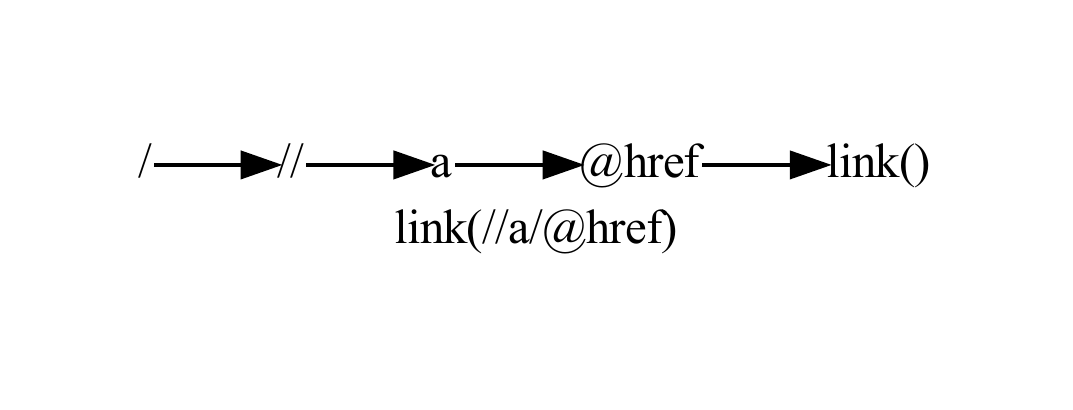}

\caption{Example XPath expression `link(//a/@href)` rendered as a graph.}
\end{figure}

\begin{figure}[h]

\includegraphics[width=1\textwidth,height=0.15\textheight]{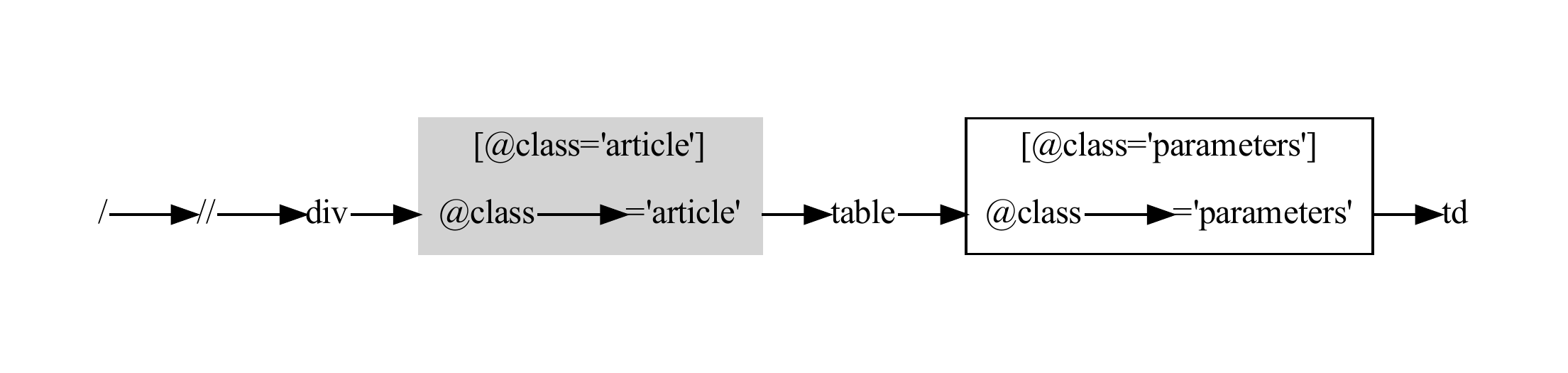}

\caption{Example XPath expression `//div[@class='article']//table[@class='parameters']/td` rendered as a graph.}
\end{figure}

In our system we only need to consider
the XPath axes that correspond to
simplest structural relations:

\begin{itemize}
\tightlist
\item
  \texttt{child::*},
\item
  \texttt{parent::**},
\item
  \texttt{preceding-sibling::*},
\item
  \texttt{following-sibling::*}.
\end{itemize}

\hypertarget{xpath-be}{%
\subsection{XPath BE}\label{xpath-be}}

Here we describe \emph{Browser
Extensions} to XPath
language{[}{\textbf{???}}{]} that allow
navigation consistent with visual
rendering and logical structure of the
web page, beside classical axes that
navigate along the tree structure of the
DOM{[}4{]} document.

\hypertarget{table-selectors}{%
\subsubsection{Table
selectors}\label{table-selectors}}

We also introduce new, semantic
selectors to XPath in order to
facilitate high-level analysis of
tables:

\begin{itemize}
\item
  within a table, we add \texttt{row::*}
  selector that selects all
  \texttt{\textless{}td\textgreater{}\textquotesingle{}s\ within\ a\ single\ table\ row\ \ \ (technically\ same\ as}../td\texttt{for\ within\ a}td\texttt{,\ and}td\texttt{from}tr`
  node)
\item
  \texttt{column::*} means all the
  entries in the same table column as a
  given \texttt{th} or \texttt{td}
  element. For example, to find all
  \texttt{td} elements within the same
  column, as \texttt{th} element with a
  text node containing exact String
  \texttt{"Address"}, we write:
  ``\texttt{//table/thead/th{[}text()=="Address"{]}/column::td}"
\end{itemize}

\hypertarget{visual-relations}{%
\subsubsection{Visual
relations}\label{visual-relations}}

We can use
\href{https://www.w3.org/TR/webdriver/\#dfn-getboundingclientrect}{WebDriver
interface}{[}13{]} to find the bounding
box of each node in HTML DOM, and thus
allow querying visual relations between
tree nodes.

We propose to translate all visual axes
into range queries in order to implement
the new visual axes:

\begin{itemize}
\tightlist
\item
  \texttt{contained-in::*} -- whether
  DOM node is encompassed within another
  node
\item
  \texttt{overlaps::*} -- whether a DOM
  node is visually overlapping with
  another node
\item
  \texttt{right::*}, \texttt{left::*},
  \texttt{up::*}, \texttt{down::*} --
  whether DOM nodes bounding box is
  strictly in the given direction on the
  web page
\end{itemize}

Additionally, we propose to query for
current font name and family used to
visualise text within the node in order
to pick the visual cues:

\begin{itemize}
\tightlist
\item
  \texttt{font-family()}
\item
  \texttt{font-style()}
\end{itemize}

\hypertarget{image-tagging}{%
\subsubsection{Image
tagging}\label{image-tagging}}

Given the rise of automatic image
processing, we also want to match on the
tags assigned to images by AI
algorithms{[}5, 6, 11, 12{]}. We propose
the syntax:

\begin{itemize}
\tightlist
\item
  \texttt{imagetags(\$nn,\ .//img)} for
  all tags inferred for each image,
\item
  and
  \texttt{imagetag(\$nn,\ .//img,\ "bottle")}
  for just checking if an image contains
  the bottle.
\end{itemize}

Here \texttt{\$nn} variable should be
defined by XPath processor (like
JavaScript XPath engine) to define the
neural network model for image tagging.

\hypertarget{lazy-web-crawling}{%
\subsubsection{Lazy web
crawling}\label{lazy-web-crawling}}

Since our goal is to find the data on a
set of web pages (for example API
portal), we need to add facilities to
deal with these. We propose the addition
of a new axis \texttt{link::*} that
connects any attribute or text
containing URL with the link target.

Thus
\texttt{//a/@href/link::*}\footnote{Or
  \texttt{//element::a/attribute::href/link::*}}
would follow any (and all) \texttt{href}
links in the document. We note that
following \texttt{link::*} or
constructing a URL with \texttt{link()}
function would be the only way to reach
elements on an alternate web page.

\hypertarget{inferxpath-meta-language}{%
\section{InferXPath
meta-language}\label{inferxpath-meta-language}}

On top of the declarative language
\textbf{XPath with browser extensions}
(\textbf{XPath-BE}) we build language
that operates on node sets \textbf{and}
the \textbf{XPath-BE} expressions
themselves.

\hypertarget{semantic-sets}{%
\subsection{Semantic
sets}\label{semantic-sets}}

Node sets are first created by semantic
set finding. Typical semantic sets
recognised on pure textual basis may be
numbers, currency amounts, HTTP transfer
methods, or JSON expressions. After
preprocessing to identify these, we
match each semantic value with the root
node of its subtree on the web page.
Thus we receive several starting
semantic sets. We can also subdivide
these semantic sets into subsets by the
DOM tree structure on the page. Another
way of describing semantic nodes perhaps
just offering an XPath that matches
them, for example \texttt{//dd} will
match all definiens nodes of definition
lists within a document.

\hypertarget{relations}{%
\subsection{Relations}\label{relations}}

When analysing relations between
elements on a web page, it is common to
use XPath expressions travelling from
the current node to another. Indeed
numerous tools facilitate this process.
We build on this to provide a language
that takes two sets of nodes and finds
an XPath that goes from nodes in one set
to all the nodes in another set.

The basic idea is that for any two sets
of values \texttt{A} and \texttt{B}, we
need a function \texttt{allPaths(A,B)}
that finds the simplest path description
between two sets of values. To simplify
the matter, we can either use unique
node identifiers for \texttt{A} and
\texttt{B} sets, or XPath expression
that finds these sets.

\hypertarget{path-reconstruction-operators}{%
\subsubsection{Path reconstruction
operators}\label{path-reconstruction-operators}}

\begin{enumerate}
\def\labelenumi{\arabic{enumi}.}
\tightlist
\item
  For a given answer set, finding a set
  of XPaths that refer to it:

  \begin{itemize}
  \tightlist
  \item
    \texttt{allPaths(//*{[}contains(text(),\ "GET"){]})}
    - enumerates XPaths that refer to a
    given set of elements for a given
    set of web pages
  \item
    \texttt{samplePaths(//*{[}contains(text(),\ "The\ first\ example"){]})}
    - enumerates XPaths that have a
    given element somewhere in the
    answer set
  \end{itemize}
\item
  We order the XPaths by the preference
  function of Path Complexity described
  in the next section
\item
  We also provide operations that filter
  out all undesired paths:

  \begin{itemize}
  \tightlist
  \item
    withPrefix(//h1) - all paths that
    start with a given path or from a
    given elements (thus having this
    answer set at the start of the
    prefix)
  \item
    \texttt{byAxis::attribute} - path
    that starts with given axis (which
    can be combined:
    \texttt{byAxis::attribute\textbar{}byAxis::child-of})
  \end{itemize}
\item
  We also defined the postprocessing
  functions:

  \begin{itemize}
  \tightlist
  \item
    \texttt{dropPrefix(//h1)} - drop
    prefix that has a given answer set
    as a result
  \end{itemize}
\item
  We also define a function that forces
  searching for a path only within a
  subtree of a web page:

  \begin{itemize}
  \tightlist
  \item
    \texttt{withinPrefix(/body/div{[}@id=\textquotesingle{}content\textquotesingle{}{]})}
    would constrain our search to within
    the
    \texttt{\textless{}div/\textgreater{}}
    element that contains the main
    content of the document.
  \end{itemize}
\end{enumerate}

Note that the key to implementing
efficient path reconstruction is
allowing the combination of XPath
complexity preference function and path
reconstruction operators on whole sets
of possible XPaths, and thus quickly
narrowing the search tree to the subset
of operators. When returning result we
compute these lazily and only return
first N results\footnote{This is easier
  to implement than making sure the
  process finishes quickly or at all.}

\hypertarget{path-complexity-preference-function}{%
\subsubsection{Path complexity
preference
function}\label{path-complexity-preference-function}}

We present the following preferences:

\begin{itemize}
\tightlist
\item
  We prefer the axes in the following
  order:

  \begin{enumerate}
  \def\labelenumi{\arabic{enumi}.}
  \tightlist
  \item
    \texttt{child::*} or
    \texttt{attribute::*}
  \item
    \texttt{following-sibling::*} and
    \texttt{preceeding-sibling::*}
  \item
    \texttt{descendant::*} Note that
    this preference function can be
    reformulated simpler: \emph{we
    always prefer axes going downward in
    the tree (stratification), and then
    those axes that have fewer direct
    connections (limiting search
    options)} One can convert it into an
    algorithm by only generating
    preferred expressions
    first\footnote{Parts of this module
      will likely be implemented by
      Migamake, if it proves to be too
      difficult.}
  \end{enumerate}
\item
  In case that result occurs on a single
  level of the tree, we prefer paths
  that do not include axes that may jump
  multiple levels
  (\texttt{following::*})
\item
  Finally we prefer shorter paths over
  the longer ones.
\end{itemize}

Now please note that giving the desired
result as answer set means that we can
enumerate connecting XPaths in reverse
order:

\begin{itemize}
\tightlist
\item
  first attempting to go upwards,
\item
  then going backwards in the tree by
  the axes that give the fewest options
\end{itemize}

\hypertarget{discussion}{%
\section*{Discussion}\label{discussion}}
\addcontentsline{toc}{section}{Discussion}

We propose the expansion of XPath with
both new relations between nodes that
correspond to the rich semantics of the
HTML Document Object Model, and
meta-level functions that allow us to
\emph{infer} the most straightforward
possible schema for the data structure
of the web page. This is with an eye on
applications similar to translation of
REST API HTML documentation web pages
into relation tables{[}1{]}.

We also note that the finding of optimal
XPath for a given application is a
common activity of website programmers
and scrapers alike, so the declarative
metalanguage for inferring XPath-BE
expressions allows for wide set of
future applications.

This meta-level reasoning is facilitated
by node-set semantics of classical
XPath, and allows us to reformulate
website schema discovery as finding
simplest XPath expressions that allow us
to navigate between different sets of
nodes.

This work-in-progress report may also be
interesting for researchers of XPath and
tree-processing languages that consider
future extensions of these that expand
the range of practical applications. We
observe that these extensions greatly
increase ergonomics, and are hard to
emulate with purely
tree-structure-oriented processing which
is used by conventional XPath and CSS
selectors.

\hypertarget{acknowledgments}{%
\section*{Acknowledgments}\label{acknowledgments}}
\addcontentsline{toc}{section}{Acknowledgments}

Implementation of this proposal is a
joint work in progress by Migamake and
KikaiTech to automate laborious scraping
tasks.

\hypertarget{references}{%
\section*{Bibliography}\label{references}}
\addcontentsline{toc}{section}{Bibliography}

\hypertarget{refs}{}
\begin{cslreferences}
\leavevmode\hypertarget{ref-haskell.love}{}%
{[}1{]} Agile generation of cloud api
bindings with haskell: 2020.
\emph{\url{https://www.youtu.be/watch?v=KY27LsV11Rg}}.
Accessed: 2020-08-10.

\leavevmode\hypertarget{ref-XPath}{}%
{[}2{]} Clark, J. and (eds.), S.D. 1999.
\emph{XML Path Language (XPath) Version
1.0}. W3C.

\leavevmode\hypertarget{ref-automated-web-extraction}{}%
{[}3{]} Crescenzi, V. and Mecca, G.
2004. Automatic information extraction
from large websites. \emph{J. ACM}. 51,
(2004), 731--779.

\leavevmode\hypertarget{ref-dom}{}%
{[}4{]} DOM, Living Standard:
\emph{\url{https://dom.spec.whatwg.org/}}.

\leavevmode\hypertarget{ref-deep-residual-image-recognition}{}%
{[}5{]} He, K. et al. 2016. Deep
residual learning for image recognition.
\emph{2016 IEEE Conference on Computer
Vision and Pattern Recognition (CVPR)}.
(2016), 770--778.

\leavevmode\hypertarget{ref-imagenet}{}%
{[}6{]} Krizhevsky, A. et al. 2017.
ImageNet classification with deep
convolutional neural networks.
\emph{CACM} (2017).

\leavevmode\hypertarget{ref-semistructured-ml}{}%
{[}7{]} Lemay, A. 2018. Machine learning
techniques for semistructured data.
(2018).

\leavevmode\hypertarget{ref-semistructured-schema-extraction}{}%
{[}8{]} Nestorov, S. et al. 1998.
Extracting schema from semistructured
data. \emph{Proceedings of the 1998 ACM
SIGMOD International Conference on
Management of Data} (New York, NY, USA,
1998), 295--306.

\leavevmode\hypertarget{ref-semistructured}{}%
{[}9{]} Quass, D. et al. 1997. Querying
semistructured heterogeneous
information. \emph{Journal of Systems
Integration}. 7, (1997), 381--407.

\leavevmode\hypertarget{ref-semistructured-ilp}{}%
{[}10{]} Ramakrishnan, G. et al. 2008.
Using ilp to construct features for
information extraction from
semi-structured text. \emph{Inductive
logic programming} (Berlin, Heidelberg,
2008), 211--224.

\leavevmode\hypertarget{ref-large-scale-image-recognition}{}%
{[}11{]} Simonyan, K. and Zisserman, A.
2015. Very deep convolutional networks
for large-scale image recognition.
\emph{CoRR}. abs/1409.1556, (2015).

\leavevmode\hypertarget{ref-inception}{}%
{[}12{]} Szegedy, C. et al. 2017.
Inception-v4, inception-resnet and the
impact of residual connections on
learning. \emph{AAAI} (2017).

\leavevmode\hypertarget{ref-webdriver}{}%
{[}13{]} WebDriver, W3C Recommendation:
\emph{\url{https://www.w3.org/TR/webdriver1/}}.
\end{cslreferences}

\bibliography{infer-xpath.bib}

\end{document}